 \documentclass[10pt,preprint]{aastex}  
 \usepackage{natbib}
\def\ang{\AA}

\def\gapprox{\lower.4ex\hbox{$\;\buildrel >\over{\scriptstyle\sim}\;$}}
\def\lapprox{\lower.4ex\hbox{$\;\buildrel <\over{\scriptstyle\sim}\;$}}

\shortauthors{ASCHWANDEN ET AL. 2012}
\shorttitle{GOES Flare Statistics}

\begin{document}

\title{		Automated Solar Flare Statistics in Soft X-rays 
		over 37 Years of GOES Observations - The Invariance of 
		Self-Organized Criticality during Three Solar Cycles
		\footnote{Manuscript version, 2012-May-28}}

\author{        Markus J. Aschwanden and Samuel L. Freeland}

\affil{         Lockheed Martin Advanced Technology Center,
                Solar \& Astrophysics Laboratory,
                Org. ADBS, Bldg.252,
                3251 Hanover St.,
                Palo Alto, CA 94304, USA;
                e-mail: aschwanden@lmsal.com}

\begin{abstract}
We analyzed the soft X-ray light curves from the {\sl Geostationary Operational
Environmental Satellites (GOES)} over the last 37 years (1975-2011) and
measured with an automated flare detection algorithm over 300,000 solar
flare events (amounting to $\approx 5$ times higher sensitivity than the NOAA 
flare catalog). We find a powerlaw slope of $\alpha_F=1.98\pm0.11$ for the
(background-subtracted) soft X-ray peak fluxes that is invariant through
three solar cycles and agrees with the theoretical prediction $\alpha_F=2.0$
of the {\sl fractal-diffusive self-organized criticality (FD-SOC)} model.
For the soft X-ray flare rise times we find a powerlaw slope of $\alpha_T
=2.02\pm0.04$ during solar cycle minima years, which is also consistent
with the prediction $\alpha_T=2.0$ of the FD-SOC model. During solar cycle
maxima years, the powerlaw slope is steeper in the range of $\alpha_T \approx 2.0-5.0$, 
which can be modeled by a solar cycle-dependent flare pile-up bias effect. 
These results corroborate the FD-SOC model, which predicts a powerlaw slope
of $\alpha_E=1.5$ for flare energies and thus rules out significant nanoflare
heating. While the FD-SOC model predicts the probability distribution functions
of spatio-temporal scaling laws of nonlinear energy dissipation processes,
additional physical models are needed to derive the scaling laws between the
geometric SOC parameters and the observed emissivity in different
wavelength regimes, as we derive here for soft X-ray emission. The FD-SOC model 
yields also statistical probabilities for solar flare forecasting. 
\end{abstract}


\section{          	INTRODUCTION  			}

There are a number of intriguing questions that have been raised about 
statistics of solar flares: 
(1) Why do they show the ubiquitous powerlaws in their
occurrence frequency distributions? (2) Do we understand the numerical
values of the powerlaw slopes?  (3) Do the powerlaw slopes vary with 
the solar cycle?  (4) Is the solar corona heated by nanoflares ?
(5) Can the flare statistics be explained 
in terms of the self-organized criticality (SOC) concept? (6) What are the
consequences of SOC models on our physical understanding of solar 
flares? (7) How can we improve statistical solar flare prediction?
(8) What are the largest expected solar flares in history and future?
In this paper we address these questions by conducting statistics of
solar flares using the longest uniform dataset we have available,
namely the soft X-ray light curves from the {\sl Geostationary Operational
Environmental Satellites (GOES)}, which cover over 37 years during the
period of 1974-2012.

The GOES program consists of a series of geostationary satellites (orbiting
the Earth at a height of 35,790 km), which overlap in time so that there
are always one to three spacecraft present and guarantee an essentially
uninterrupted time series of solar soft X-ray fluxes, besides continuous
meteorological observations of the Earth. The GOES-1 satellite was 
launched on October 16, 1974, and GOES-2 and GOES-3 followed in 1977 and
1978. In the meantime the series continued up to GOES-15, launched on
March 4, 2010, while the future satellites GOES-R and GOES-S (with soft X-ray
imaging capabilities) are in the queue for a launch in 2015 and 2017,
respectively. Operational and technical details of GOES satellites can
be gleaned from Grubb (1975), Donnelly et al.~(1977), Bouwer et al.~(1982), 
Thomas et al.~(1985), Kahler and Kreplin (1991), Garcia (1994), Hill (2005), 
Pizzo et al.~(2005), Shing et al. (1999), White et al.~(2005), Neupert (2011),
the {\sl National Oceanographic and Atmospheric Administration
(NOAA)} website {\sl http://www.oso.noaa.gov/goes/}, or the NASA website
{\sl http://goespoes.gsfc.nasa.gov/project/index.html}. For our study 
we are concerned with the soft X-ray light curves, which are recorded in 
two energy channels, i.e., in the softer energy range of 1-8 \ang\ and 
in the harder energy range of 0.5-4 \ang. Both light curves are available 
with a cadence of 3 s, and with a cadence of 2 s after 2009 Dec 1. 

Solar flare statistics can be generated by detection of sudden impulsive
increases of the soft X-ray intensity in light curves. NOAA publishes
solar flare catalogs that are issued on a daily basis. Here we develop
an automated flare detection algorithm that allows us to analyze the
entire 37-year time series in an objective way without human subjectivity.
We are interested in the mathematical function of the probability 
distributions of solar flare parameters. It has been established earlier
that the size distribution of soft X-ray peak fluxes of solar flares
have a powerlaw-like form, with a powerlaw slope in the range of
$\alpha_P \approx 1.6-2.1$
(Hudson et al. 1969, Drake et al. 1971, Shimizu 1995, Lee et al. 1995,  
Feldman et al. 1997, Shimojo and Shibata 1999, Veronig et al. 2002a,b, 
Yashiro et al. 2006). In this study we will measure this parameter with
the largest available statistics, including over 300,000 detected solar
flare events, and will study its variation during three solar cycles.
Further we will investigate measurement uncertainties of previously
analyzed distributions and compare them with the results from this study
in order to establish the most accurate value.

The appearance of powerlaw-like size distribution functions has generally
been linked to the theory of {self-organized criticality (SOC)} (or
alternatively to turbulence), which predicts scale-free powerlaw 
distributions (Bak et al.~1987, 1988). A prominent paradigm
of the SOC concept is a sandpile that produces a scale-free distribution
of avalanche sizes, once it reaches a critical slope and is slowly
driven by random input of dropped sand grains. This concept was also
applied to the magnetically driven solar corona, which releases 
intermittent amounts of nonpotential magnetic energies during flares
(Lu and Hamilton 1991). Many applications of SOC processes can be found
from geophysics to astrophysics, all the way to financial and social
systems. The topic of SOC is reviewed in recent reviews,
textbooks, and monographs (e.g., Bak 1996; Jensen 1998; Turcotte 1999;
Charbonneau et al.~2001; Hergarten 2002; Sornette 2004; Aschwanden 2011a;
Crosby 2011; Pruessner 2012). In this study we go beyond the mere
establishment of powerlaw indices, but try also to understand the
quantitative values of the observed powerlaw slopes and their variation
with the solar cycle. We interpret the observations in terms of a recently
published theoretical model based on a statistical fractal-diffusive
avalanche model in a slowly-driven self-organized criticality system
(Aschwanden 2012). The application of this model allows us also to
provide a quantitative framework for statistical flare predictions, 
which includes also probabilities for the most extreme space weather events.

The content of this paper is as follows: Section 2 presents the
data analysis and results, in Section 3 we apply the SOC theoretical model,
in Section 4 we compare the results with previous observations, and in
Section 5 we summarize the conclusions.

\section{          	DATA ANALYSIS 			}

\subsection{		Dataset 			}

Our basic data input is the 37-year time series of complete years (from
1975 to 2011) of the GOES softer X-ray channel in the 1-8 \ang\ wavelength
range. We choose the softer energy channel (1-8 \ang\ = 1.5-12 keV) over 
the harder energy range (0.5-4 \ang\ = 3-25 keV) because of its higher
sensitivity to smaller flares and less data noise. The data are available
in the {\sl Solar Software (SSW)} (Freeland and Handy 1998), 
originally provided by NOAA, and can
be read with IDL software (using the procedure {\sl $RD\_GXD.PRO$}). The data
are returned with time tick marks $t_i, i=1,..,N$ for the 3-s intervals 
and flux values $f_i^{\lambda}, i=1,...,N$ in physical units of $(W m^{-2})$,
for both wavelength ranges $\lambda=1-8$ \ang\ and $\lambda=0.5-4$ \ang .
The logarithmic flux values are also labeled with letters (A, B, C, M, X-class),
which denote the order of magnitude of the peak flux on a logarithmic scale
($A=10^{-8}, B=10^{-7}, C=10^{-6}, M=10^{-5}, X=10^{-4}$ $W m^{-2}$),
subdivided with an additional digit (e.g., an X2 class flare has a flux
of $2 \times 10^{-4}$ $W m^{-2}$).

Flare catalogs are issued by NOAA, which provide the peak fluxes, start
times, peak times, and end times of flare events, but no preflare background
is provided in the flare catalogs, which is important for statistical
studies of small flares. If no preflare background is subtracted from
the peak flux, the peak flux includes not only the flare-associated flux,
but also the total soft X-ray flux of all active regions on the entire
solar disk and beyond the limb, which is larger than the flare-associated
flux for A, B and C-class flares during active phases of the solar cycle 
(Wagner 1988; Bornmann 1990; Aschwanden 1994; Veronig et al.~2004).
Therefore, proper evaluation of the preflare background requires the analysis
of the light curves, and cannot be obtained from the official NOAA flare
catalog alone. Furthermore, the GOES light curves contain occasional
data gaps (due to Earth occultation, calibration procedures, or drop-outs)
and nonsolar spikes (of instrumental, terrestrial, or magnetospheric origin),
which need to be removed to avoid false flare detections.

\subsection{	Automated Flare Detection Algorithm	}

After testing various automated flare detection algorithms at various
background soft X-ray flux levels, which range from the A-level during
the solar cycle minimum to the C-level during the solar cycle maximum, 
by optimizing
maximum sensitivity of detecting the smallest flares and 
by minimizing false detections, we arrived at the following algorithm:

\begin{enumerate}
\item{\underbar{Rebinning of data:} The intrinsic time resolution of $dt=3$ s
	is rebinned to time steps of $\Delta t=12$ s (matching the
	SDO/AIA cadence used in another study) by the median value
	during each bin with length $n_{bin}=\Delta t/dt=12/3=4$.
	Thus, a daily record has $n_{day}=86,400/3=28,800$ datapoints,
	which is downsampled to a number of $n_{bin}=86,400/12=7200$ bins.}   

\item{\underbar{Definition of minimum flare duration:} We set a minimum flare
	duration at a 1-minute time interval, i.e., $\Delta t_{min}=60$ s,
	which corresponds to a number of $\Delta N_{bin}=60/12=5$ bins.}

\item{\underbar{Definition of noise level and threshold:}
	From the mean and standard
	deviation of typical GOES fluxes during quiescent time periods we
	find a noise level of $f_{noise} \approx 2 \times 10^{-8}$ $W m^{-2}$,
	and define a corresponding threshold level of $f_{thresh}=f_{noise}$.
        The noise level of GOES-8, 9, 10 is constant
        within less than 10\% variation (http://rammb.cira.colostate.edu/
        research/
	calibration$\_$validation$\_$and$\_$visualization/goes$\_$image$\_$display/
        noise.asp).}

\item{\underbar{Elimination of data gaps:} These are identified by time
	intervals with a constant ``floor'' flux value that corresponds 
	to the minimum flux of the daily light curve.} 

\item{\underbar{Spike removal:} Single spikes are detected if the ratio of the
	maximum flux $f_{max}$ to the minimum flux $f_{min} \ge f_{noise}$
	is larger than $q_{spikes} = f_{max}/f_{min} > 10$ during a 
	time interval of $\pm \Delta t_{min}=60$ s.}

\item{\underbar{Smoothing of light curve:} 
	We smooth the rebinned light curve with a
	boxcar of $nsm=21$ time bins, which corresponds to 
	$nsm \approx 4 \Delta N_{bin}$ minimum flare durations.}

\item{\underbar{Detection of maxima and minima:} 
	We detect now all local maxima and
	minima of the smoothed light curve (consecutively in daily intervals).
	The flux maximum times $t_i$ represent candidates for flare peak times,
	and flux minimum times $t_{i-1}$ and $t_{i+1}$ represent potential
	flare start and end times.}

\item{\underbar{Detection of flare event:} Flare events are defined 
	when they fulfill the following conditions (see Fig.~1): 
	(i) The flare starts at a flux
	minimum time $t_s=t_i$, where a preflare background $f_{BG}$
	is defined from the median flux in a time interval $[t_s-\Delta 
	t_{min},t_s]$; (ii) The flare ends at the first subsequent flux 
	minimum time $t_e=t_{i+1+k}$ with $k > 0$ when the flux drops below the
	level $f_{BG}+f_{thresh}$, but before the start of a next flare
	(i.e., no overlapping flare time intervals); 
	(iii) The flare peaks at the highest
	flux value $f_p$ at time $t_p$ during the intervening
	time interval $t_s < t_p < t_e$. (iv) The background-subtracted
	peak flux is $F=f_p-f_{BG}$. 
	Obviously, during solar maximum a lot of flares start on the tail 
	of previous larger events, and thus the background flux is
        higher, which is not the case during solar minimum. The GOES
        non-flare background flux varies similarly to the sunspot number
        (Wagner 1988). - In addition we measure also the
	steepest flux derivative $(df/dt)_{max}$ during the rise time 
	$[t_s, t_p]$, which can be used as a proxi for the hard X-ray
	peak flux according to the Neupert effect (Dennis and Zarro 1993).}
\end{enumerate}

An example of automated flare detection is shown for 2012-Jan-27 in Fig.~2,
when an GOES class X1.8 flare occurred. A total of 36 flare events were
detected during this day, mostly ranking in the B and C-class level.
In this example we see that the flux of most of the small B and C-class 
flares returns to a background level of $\approx 5 \times 10^{-7}$ 
$W m^{-2}$. The duration of larger flares (e.g., C5.5, C.3.3, X1.8,
C3.5) is truncated by the start of subsequent flares. Therefore, we 
expect that the flare duration is generally underestimated at times of
high flare rates. A more correct flare duration could potentially be
derived from a flare decay model that extrapolates the beginning of the
flare decay phase all the way to the background level.

We found that this flare detection algorithm is sensitive to the smallest
flares recognizable by visual inspection, and at the same time to be
fairly robust with an estimated false-detection rate of 
$\lapprox 10^{-2} ... 10^{-3}$, based on visual flare classification.

\subsection{	Flare Rate and GOES Duty Cycle 		}

The monthly rate of detected flares is shown in Fig.~3, for all flares,
as well as for the subsets of $>C1.0$, $>M1.0$, and $>X1.0$ flares
separately. The total number of detected flares larger than a given
magnitude decreases approximately by a factor of 10 for each order of
magnitude: from 338,661 events of all flares to 35,221 $>C$-class flares,
3986 $>$M-class flares, to 248 $>$X-class flares, during a time span
of 37 years. The detected flare rate is also shown on a logarithmic
scale (Fig.~3 bottom), which shows better the proportionality of detected
flares in different magnitudes, even across the minima and maxima of the
three solar cycles. There are indeed some months during the last extended
solar cycle minimum 2008-2009 with no detected flares at all, which
corroborates also the robustness of our detection algorithm for false
events by noise coincidence. The number of detected flares per year
is listed in Table 1, ranging from a minimum of 186 events per year (2008) 
to a maximum of 18,797 events per year (1979).

Comparing the number of detected flares with the official GOES flare catalog
from NOAA (readable with procedure $GET\_GEV.PRO$ in $SSW/IDL$), we find
39,696 flare events reported by NOAA during the period of 1991-2011 
(the annual number of NOAA events is listed in Table 1, 3rd column). 
Thus our flare detection algorithm detects 9153 events per year in the
average, while the NOAA catalog lists 1804 events per year, so our algorithm
is about 5 times more sensitive. The NOAA flare detection algorithm is defined
as follows: The event starts when 4 consecutive 1-minute X-ray values have met 
all three of the following conditions: (i) All 4 values are above the B1 
threshold; (ii) All 4 values are strictly increasing; (iii) The last value 
is greater than 1.4 times the value that occurred 3 minutes earlier. 
The peak time is when the flux value reaches the next local maximum. 
The event ends when the current flux reading returns to half of the peak
value (http://www.ngdc.noaa.gov/stp/solar/solarflares.html).

We determined the effective GOES duty cycle from the number of datapoints
that we obtained from reading the GOES database with the SSW standard
routine and found an average of $94\%\pm4\%$ during the years 1978-2011,
and a lower fraction of $76\%\pm8\%$ during the first 3 years of the GOES
series (see Table 1). The actual duty cycle of GOES may be higher, because 
our value of the duty cycle is derived from the number of readable files, 
and thus the missing data may also include file reading problems, besides 
the originally missing data (due to data loss, telemetry gaps, calibration
procedures, Earth occultation, etc.). Anyway, GOES has probably the 
highest duty cycle from all solar-dedicated space missions, and thus
offers the most complete record of solar flares over the last three
solar cycles. 

\subsection{	Occurence Frequency Distributions	}

The size distribution of many solar flare parameters has been found
to be close to a powerlaw function. However, often there is a gradual
turnover at the lower end of the powerlaw range near the detection
threshold. In addition, there is sometimes an exponential drop-off
apparent at the upper end of the powerlaw range. Since the fit of
a powerlaw slope can be severely affected by the gradual turnover
at the lower end, we fit a powerlaw function with an empirical
correction term that characterizes the turnover with a transition 
to a constant function at the lower end of the distribution function, 
\begin{equation}
	N(x) dx = N_1 ( 1 + x/x_1 )^{-\alpha} dx \ .
\end{equation}
At the low end the distribution converges to a constant
$N(x \ll x_1) \mapsto N_1$, while the asymptotic limit at the
upper end turns into a pure powerlaw function, $N(x \gg x_1)
\mapsto N_1 (x/x_1)^{-\alpha}$. We call $x_1$ the ``lower bound''
or ``turnover value'' of the powerlaw range.

The powerlaw fits of the size distribution functions of the soft X-ray
flux are shown for each year from 1974 to 2012 separately in Fig.~4,
yielding powerlaw slopes in the range of $\alpha_F=1.7-2.3$, 
with a mean and standard deviation of $\alpha_F=1.98\pm0.11$. 
The individual values and uncertainties are also listed for every
year in Table 1. The annual fits show in most cases a clear turnover 
at the lower end, which is well fitted by our empirical function (Eq.~1)
and renders the value of the powerlaw slope quite stable, independently
of which bins near the turnover are included or excluded in the fit.
The accuracy of the powerlaw slope is often hampered in earlier studies
due to arbitrary definitions of the fitted range. 

In Fig.~5 we compare the powerlaw fits for the GOES peak fluxes for
different phases of the solar cycles. We select three different regimes:
(i) Years near the solar cycle maximum (defined by periods with more
than 10,000 flare events detected per year, see Table 1, second column,
which includes the periods of 1978-1982 for Cycle 21, 1988-1991 for
Cycle 22, 1998-2003 for Cycle 23, and 2011 for Cycle 24),
(ii) Years near the solar cycle minimum (defined by periods with less
than 3000 flare events detected per year, which includes the periods of
1975-1976, 1985-1986, 1995-1996, and 2007-2009), and (iii) intermediate
time intervals during the rise or decay of solar cycles. 
The powerlaw fits of these three data subsets exhibit a remarkable
constant powerlaw slope of $\alpha_F \approx 2.0$ (Fig.~5, top panel), 
so that the probability distribution functions differ only by a
variable scaling factor for the occurrence frequency as a function of time.

In Fig.~5 (middle panel) we show the same statistics for the parameter
of the maximum flux-time derivative $(df/dt)_{max}$ during the rise time
of the flare light curve, which also exhibits the same behavior as the
soft X-ray peak flux, namely a constant powerlaw slope $\alpha_{df/dt}
\approx 2.0$, with a variable scaling factor for the occurrence
frequency as a function of time. The powerlaw slopes $\alpha_{df/dt}$
are also tabulated for each year in Table 1. If we would apply the
Neupert effect, the time derivative $(df/dt)$ of the soft X-ray flux
should be a good proxi of the (smoothed) hard X-ray peak flux, which
predicts then a powerlaw slope of $\alpha_P^{pred}=\alpha_{df/dt}=2.0$.
In reality, however, hard X-ray peak rates were found to have a powerlaw
slope of $\alpha_P=1.73\pm0.07$ (Aschwanden 2011b). This discrepancy
results from a combination of two effects: (i) The simplest formulation 
of the Neupert effect in terms of an integral (Eq.~21) is oversimplified 
as we will see below (Section 3.2), and (ii) the time derivative measured 
from the {\sl smoothed} soft X-ray time profile represents an underestimate 
of the true time derivative of the unsmoothed light curve.
 
Finally, we plot also the frequency distribution of flare rise times
$t_{rise}$, which is probably a good proxy for the flare duration $T$ 
of effective energy release, as measured from non-thermal X-rays light
curves and expected from the Neupert effect. We plot the size distributions 
for the same three epochs of the solar cycle, but find
a completely different behavior (Fig.~5, bottom panel). The powerlaw
slopes are found to be steepest at the solar cycle maximum with an
average slope of $\alpha_T \approx 3.2$, while they become flatter
with a value of $\alpha_T \approx 2.3$ at the solar minimum.
The values for different years vary from $\alpha_T=1.75$ (for the
year 2008 with the most extreme solar minimum), up to $\alpha_T=5.18$
(for the year 2002, which represented the peak of the last solar Cycle 23),
as it can be seen from the list in Table 1.

\subsection{	Solar Cycle Variatiability 		}

Apparently, some soft X-ray parameters vary during the
solar cycle, while others do not. For an overview of this time-dependent
behavior we plotted the various parameters as a function of time (over the
last 37 years) in Fig.~6. The most invariant parameters are the powerlaw
slope $\alpha_F \approx 2.0$ of the soft X-ray peak flux and the 
slope $\alpha_{df/df} \approx 2.0$ of the time derivative, having almost
identical values. The largest deviation from a constant value is seen for 
the year 2009, 
the year with the second-lowest flare rate ($N_{flare}=528$), and thus
represents a statistical fluctuation due to small-number statistics.

The time variation of the (detected) soft X-ray flare rate is shown
in the bottom panel of Fig.~6 (hatched with a grey area). Comparing
with this solar cycle variation of the flare rate, we see clearly that
the other parameters plotted in Fig.~6 all exhibit a correlated
time variation, which applies to the powerlaw slope $\alpha_T$
of the flare rise time $T$ (Fig.~5, 3rd panel), the
turnover value $P_1$ of the peak count rate $P$ (corresponding to the
symbol $x_1$ in Eq.~1) (Fig.~5, 4th panel), the turnover value
$(df/dt)_1$ of the time derivative $(df/dt)$ (Fig.~5, 5th panel),  
and the turnover value $T_1$ of the rise time $T$
(Fig.~5, 6th panel). In all these parameters, the turnover value
seems to fluctuate in synchronization with the flare rate. This implies 
that the degree of undersampling of weak flares becomes more severe for
times with higher flare rates, especially during the solar cycle
maxima. This behavior is apparently a feature of our automated
flare detection algorithm, which detects a smaller number of flares
during periods of high activity, either because the flux contrast for
small flares is relatively weaker on top of a light curve of a large 
flare, compared with the background noise level during quiescent times.
In addition, the rule that each flare is separated in time and does not
allow for temporal overlap (of rise times), may suppress the recording
of smaller flares during the evolution of a larger flare. Actually,
we will demonstrate in the following theoretical section that this
effect can be simply modeled and predicted as a function of the 
instantaneous flare rate (Section 3.3). 

\section{		THEORETICAL MODEL		}

We apply now the statistical fractal-diffusive (FD-SOC) avalanche model 
of a slowly-driven self-organized criticality (SOC) system, which is derived in
general form for all 3 Euclidean dimensions $S=1,2,3$ and tested with
cellular automaton simulations in a recent paper (Aschwanden 2012). 
For our application to soft X-ray data from solar flares here we consider 
only the 3D case ($S=3$) and add a simple physical model of the 
soft X-ray (thermal) radiation mechanism (Section 3.2). 

\subsection{	The Fractal-Diffusive Avalanche SOC Model }

The fractal-diffusive SOC model is a universal (physics-free) analytical
model that describes the statistical time evolution and occurrence
frequency distribution function of SOC processes. It is based on four
fundamental assumptions: (1) A SOC avalanche grows spatially like a
diffusive process; (2) The spatial volume of the instantaneous energy 
dissipation rate is fractal; (3) The time-averaged fractal dimension 
is the mean of the minimum dimension $D_{S,min}\approx 1$ (for a sparse
SOC avalanche) and the maximum dimension $D_{S,max}=S$ (given by the Euclidean 
space); and (4) The occurrence frequency distribution of length scales
is reciprocal to the size $L$ of spatial scales, i.e., $N(L) \propto L^{-S}$
in Euclidean space with dimension $S$. We will discuss these assumptions
in more detail in the following. 

The first assumption of a diffusive process is based on numerical
simulations of cellular automaton models. A SOC avalanche propagates
in a cellular automaton model by next-neighbor interactions in a critical
state, where energy dissipation propagates only to the next neighbor cells 
(in a S-dimensional lattice grid) that are above a critical threshold.
This mathematical rule that describes the entire dynamics and evolution
of a SOC avalanche is very simple for a single time step, but leads to
extremely complex spatial patterns after a finite number of time steps.
For a visualization of a large number of such complex spatial patterns
generated by a simple iterative mathematical redistribution rule see,
for instance, the book ``A New Kind of Science'' by Wolfram (2002).
The complexity of these spatial patterns can fortunately be characterized
with a single number, the fractal dimension $D_S$. If one monitors the
time evolution of a spatial pattern of a SOC avalanche in a cellular
automaton model, one finds that the length scale $x(t)$ evolves with time
approximately with a diffusive scaling (Section 2.1 in Aschwanden 2012),
\begin{equation}
	x(t) \propto t^{1/2} \ ,
\end{equation}
which leads to a scaling law between the avalanche sizes $L=x(t=T)$
and time durations $T$ of SOC avalanches,
\begin{equation}
	L \propto T^{1/2} \ .
\end{equation}

The second assumption of a fractal pattern of the instantaneous energy
rate is also based on tests with cellular automaton simulations (see
Fig.~2 in Aschwanden 2012). The fractal dimension is essentially a
simplified parameter that describes the ``micro-roughness'', ``graininess'',
or inhomogeneity of critical nodes in a lattice grid in the state of 
self-organized criticality. Of course, such a single number is a gross
over-simplification of a complex system with a large number of degrees of
freedom, but the numerical simulations confirm that it reproduces the
correct scaling law between the instantaneous energy dissipation volume
$V_S(t)$ and the spatial scale $x(t)$ of SOC avalanches,
\begin{equation}
	V_S(t) \propto x^{D_S} \ ,
\end{equation}
which leads also to a statistical scaling law between avalanche
volumes $V$ and spatial scales $L$ or durations $T$ (with Eq.~3) of
SOC avalanches,
\begin{equation}
	V_S \propto L^{D_S} \propto T^{D_S/2} \ .
\end{equation}

The third assumption of the mean fractal dimension has also been
confirmed by numerical simulations of cellular automaton SOC processes
in all three dimensions $S=1,2,3$ (Aschwanden 2012), but it can also 
be understood by the
following plausibility argument. The sparsest SOC avalanche that 
propagates by next-neighbor interactions is the one the spreads only
in one spatial dimension, and thus yields an estimate of the minimum
fractal dimension of $D_{S,min} \approx 1$, while the largest SOC avalanche
is almost space-filling and has a volume that scales with the Euclidean
dimension, $D_{S,max}=S$. Combining these two extreme cases, we
can estimate a time-averaged fractal dimension from the arithmetic mean,
\begin{equation}
	<D_S> \approx {D_{S,min}+D_{S,max} \over 2} = {1 + S \over 2} \ ,
\end{equation}
which yields a mean fractal dimension of $<D_3>=(1+3)/2=2.0$ for the
3D case $(S=3)$.

The last assumption of the size distribution is a probability argument.
The system size $L_{sys}$ of a SOC system represents an upper limit of 
spatial scales $L$ for SOC avalanches, i.e., $L \le L_{sys}$. For the
3D-case, the volumes $V$ of individual avalanches are also bound by
the volume $V_{sys}$ of the system size, i.e., $V = L^3 \le V_{sys} =
L_{sys}^3$. If the entire system is in a critical state, SOC avalanches
can be produced everywhere in the system, and the probability $N(L)$ for a
fixed avalanche size L with volume V is simply reciprocal to the size,
i.e.,
\begin{equation}
	N(V) \propto {V \over V_{sys}} \propto V^{-1} \ ,
\end{equation}
which is equivalent to
\begin{equation}
	N(L) \propto {L^3 \over L_{sys}^3} \propto L^{-3} \ .
\end{equation}

Based on this model we simulate now an example of a time evolution of 
SOC parameters for the 3D case, as shown in Fig.~7. We start with
a mean fractal dimension $<D_3>=2.0$, and simulate the time evolution
of the fractal dimension $D_3(t)$ by simulating fluctuations with a
relatively small amplitude (with a standard deviation of $\sigma_D=0.15$) 
by a random generator (Fig.~7 top panel). The time evolution of the
instantaneous energy dissipation rate $f(t)$ in classical SOC systems
(with a constant mean dissipated energy quantum $<\Delta E>$ per unstable
lattice node) is then proportional to the instantaneous volume $V_S(t)$, 
which yields the following function (with Eq.~4),
\begin{equation}
	f(t) ={de(t) \over dt} \propto <\Delta E> V_S(t)
	=<\Delta E> x(t)^{D_S} 
	=<\Delta E> t^{D_S/2} \ .
\end{equation} 
In the 3D case with $D_S=2$, we expect than the proportionality 
$f(t) \propto t^{1.0}$, as shown in Fig.~7 (second panel). 

The statistical peak value $p(t)$ of the energy dissipation rate
after time $t$ can be estimated from the largest possible avalanches,
which have an almost space-filling dimension $D_S \lapprox S$, and
thus would be expected to scale as $p(t) \propto t^{S/2} \propto t^{1.5}$
(indicated with a dashed curve in Fig.~7, second panel.

The evolution of the total dissipated energy $e(t)$ after time $t$
is simply the time integral, for which we expect
\begin{equation}
	e(t) = \int_0^t {de(\tau) \over d\tau} d\tau
	\propto \int_0^t \tau^{D_S/2} \propto t^{(1+D_S/2)} \ ,
\end{equation}
which yields the function $e(t) \propto t^2$ for the 3D case
(Fig.~7, third panel). These time evolutions apply to every classical
SOC model and can be derived from purely statistical probability
arguments, and thus are physics-free. For application to real-world
data we have to understand the physical nature of the observables,
before we can relate the measured distributions to the statistical
SOC theory, which we undertake in the next Section.

\subsection{	SOC Application to Solar Flare Soft X-ray Data	}

SOC theory has been applied to different wavelength regimes of
solar flare observations, such as to gamma-rays, hard X-rays,
soft X-rays, and extreme ultra-violet (EUV). A comprehensive
review of such studies is given in Section 7 of Aschwanden (2011a).
However, since each wavelength range represents a different physical
radiation mechanism, we have to combine now the physics of
observables with the (physics-free) SOC statistics.

Soft X-ray emission during solar flares is generally believed to result
from thermal free-free and free-bound radiation of plasma that is
heated in the chromosphere by precipitation of non-thermal electrons
and ions, and which subsequently flows up into coronal flare (or
post-flare) loops, a process called ``the chromospheric evaporation
process'' (for a review see, e.g., Aschwanden 2004).
Therefore, we can consider the flare-driven chromospheric 
heating rate as the instantaneous energy dissipation process of a
SOC avalanche, as shown in the simulated function $f(t)$ in Fig.~7
(second panel). The heated plasma, while it fills the coronal flare
loops, loses energy by thermal conduction and by radiation of
soft X-ray and EUV photons, which generally can be characterized
by an exponential decay function after an impulsive heating spike.
In Fig.~7 (bottom) we mimic such a soft X-ray radiation light curve
by convolving the instantaneous energy dissipation rate $f(t)$
(Fig.~7, second panel) with an exponentially decaying radiation
function (with an e-folding time constant of $\tau_{decay})$,
\begin{equation}
	f_{sxr}(t) = \int_{-\infty}^t f(t) \exp{[-{(t-t')}/\tau_{decay}]} 
	dt' \ ,
\end{equation}
which shows also a time dependence that follows approximately
\begin{equation}
	f_{sxr}(t) \propto f(t) \propto t^{1.0} \ ,
\end{equation}
because the convolution with an exponential function with a constant 
e-folding time constant acts like a constant multiplier. 
In the limit of infinitely long decay times $(\tau_{decay} \mapsto \infty)$,
our convolution function (Eq.~11) turns into a time integral of the
heating function $f(t)$, which is also known as Neupert effect
(Dennis and Zarro 1993; Dennis et al.~2003),
where the heating function is identified with the non-thermal hard X-ray
emission and the time integral with the soft X-ray emission. However,
such an approximation would predict different values for the powerlaw
slopes of hard X-ray and soft X-ray peak fluxes, which was found to
contradict the data (Lee et al.~1995). Hence, we consider the
formulation of the Neupert effect in terms of a convolution with a
finite cooling time (Eq.~11) as a more accurate representation than
the time intergral formulation (with an infinite cooling time).

We can now calculate the occurrence frequency distributions.
The total duration $T$ of energy release corresponds essentially
to the rise time $t_{rise}$ of the soft X-ray flux, because the
decay phase of a soft X-ray flare light curve is dominated by
conductive and radiative loss, rather than by continued heating input.
Thus, starting from the size distribution of length scales,
$N(L) \propto L^{-S}$ (Eq.~8), we can derive the size distribution
of rise times $T=t_{rise}$ by substituting the variable $T$
for $L$ in the distribution $N(L)$, using the diffusive scaling
$L(T) \propto T^{1/2}$ (Eq.~3), with the derivative $dL/dT=T^{-1/2}$,
\begin{equation}
        N(T) dT = N(L[T]) \left| {dL \over dT} \right| dT
        \propto T^{-[(1+S)/2]} \ dT = T^{-2} dT \ ,
\end{equation}
yielding $N(T) \propto T^{-2}$ for $S=3$. 
Subsequently we can derive the size distribution $N(F)$ of the
energy dissipation rate $F=f(t=T)$,
using the relationship $F(T) \propto T^{D_S/2}$ (Eq.~9),
\begin{equation}
        N(F) dF = N(T[F]) \left| {dT \over dF} \right| dF
        \propto F^{-[1+(S-1)/D_S]} = F^{-2} \ dF \ .
\end{equation}
Thanks to the proportionality between the functions $f(t)$ and $f_{sxr}(t)$
(Eq.~12), 
we have also a proportionality between the statistical expectation values 
$F=f(t=t)$ and $F_{sxr}=f_{sxr}(t)$, and thus a size distribution with an
identical powerlaw index,
\begin{equation}
        N(F_{sxr}) dF_{sxr} = \propto F_{sxr}^{-[1+(S-1)/D_S]} 
	= F_{sxr}^{-2} \ dF_{sxr} \ .
\end{equation}
Thus, our theory predicts a powerlaw slope of $\alpha_F=2.0$ for the
soft X-ray fluxes (Eq.~15), and an identical slope of $\alpha_T=2.0$ for the
rise time $t_{rise}=T$ (Eq.~13). As we can see in the observations shown in 
Fig.~6, the powerlaw slopes of the soft X-ray fluxes has an annually averaged
powerlaw slope of $\alpha_F=1.98\pm0.11$, which is moreover invariant
over a time span of 37 years, and thus is fully consistent with our
theoretical prediction of $\alpha_F=2.0$. The powerlaw slopes of
soft X-ray rise times have a minimum value of $\alpha_T \gapprox 2.0$
during the solar cycle minima (Fig.~6, third panel), but deviate up
to $\alpha_T \lapprox 5.0$ during solar cycle maxima. Apparently there
is a solar cycle effect that modifies the theoretically predicted value,
which we interpret as a measurement bias in the next Section. 

\subsection{	Solar Cycle Dependence of Flare Pile-Up Bias}

Our theory predicts an occurrence frequency distribution of $N(T) \propto
T^{-2}$ for flare durations (or rise times in soft X-ray light curves)
in the slowly-driven limit where subsequent avalanches do not overlap
in time. Also, our automated flare detection algorithm does not allow for
overlapping flare time intervals. However, it is conceivable that the
flare rate is so high during the most active times of the solar cycle
that multiple flares overlap each other, which violates the slowly-driven
condition and leads to an underestimate of the flare durations
(because of our rule that a flare has to end before the next 
flare starts), especially during large and long flares, a bias that 
leads to a steepening of the powerlaw slope. Indeed we observe
powerlaws that are steeper up to a value of $\alpha_T=5.18\pm0.55$ 
(in the year 2002 near the maximum of the last solar cycle). Thus we
try to estimate this effect, which we call ``flare pile-up bias''.

The average waiting time between two subsequent flares per year 
(with a duration of $365.25 \times 86,400$ s = $3.2 \times 10^7$ s) is
\begin{equation}
	<\Delta t_{wait}> = {3.2 \times 10^7 s \over N_{flare}} \ ,
\end{equation}
which varies between 
$<\Delta t_{wait}> = {3.2 \times 10^7 s / 18,797} \approx 1700$ s
$\approx 28$ min (in the year 1979) to 
$<\Delta t_{wait}> = {3.2 \times 10^7 s / 186} \approx 170,000$ s
$\approx 2$ days (in the year 2008). The latter year is sufficiently low to
fulfill the slowly-driven condition, because the flares are so sparse
that they never overlap. However, the former year with an
average waiting time of $1.7 \times 10^3$ s is significantly shorter than the
longest measured flare rise time ($t_{rise} \lapprox 10^{4.5}$ s)
and thus underestimates the longer flare durations. We estimate the
effect of this flare pile-up bias on the measurement of the powerlaw slope,
which is generally defined as,
\begin{equation}
	\alpha_T = - {\log{ (N_2/N_1) } \over \log{ (T_2/T_1) } } \ ,
\end{equation}
with $T_1$ and $T_2$ the lower and upper bound of the powerlaw range,
and $N_1$ and $N_2$ the corresponding number of flares at these
boundaries. In the case of flare pile-ups the upper limit $T_2$ is
limited by a time scale that approximately corresponds to the
average waiting time $<\Delta t_{wait}>$, modifying the observed
powerlaw slope $\alpha_T^{obs}$ to
\begin{equation}
	\alpha_T^{obs} = - {\log{ (N_2/N_1) } \over 
	\log{ (<\Delta t_{wait}> /T_1) } } \ .
\end{equation}
for $<\Delta t_{wait}> < T_2$.
From years with sparse flare rates fulfilling the slowly-driven
condition we find $\log{ (N_2/N_1) } \approx 5.2$, $\log{ (T_2/T_1) 
\approx 2.6}$, yielding $\alpha_T^{obs} \approx 2.0$.
Combining Eq.~(16) and (18) we obtain a time-dependent powerlaw index
that depends on the annual flare rate $N_{flare}(t)$ as,
\begin{equation}
	\alpha_T^{obs}(t) = - {\log{ (N_2/N_1) } \over 
	\log{ (3.2 \times 10^7 / N_{flare}(t) \ T_1) } } \ ,
\end{equation}
if $N_{flare} \ge 3.2 \times 10^7 s / T_2 $. We calculate this
time-dependent powerlaw index $\alpha_T^{obs}(t)$ for all years 
(1975-2011) and display the expected curve in Fig.~6 (3rd panel),
which closely mimics the powerlaw rate variation $\alpha_T^{obs}(t)$ 
inferred from
the observations. Thus, we conclude that the correct powerlaw slope
of $\alpha_T=2.0$ predicted by SOC theory can be measured during
solar cycle minima, while flare pile-up causes a steeper 
powerlaw slope during solar cycle maxima. 

\section{		DISCUSSION 			}

\subsection{	Comparison with Previous Statistics	}

The size distributions of soft X-ray peak fluxes of solar flares
have been reported from various spacecraft data, with powerlaw indices
in the range of $\alpha_P \approx 1.6-2.1$
(Hudson et al. 1969, Drake et al. 1971, Shimizu 1995, Lee et al. 1995,  
Feldman et al. 1997, Shimojo and Shibata 1999, Veronig et al. 2002a,b, 
Yashiro et al. 2006). A compilation of these studies is listed in Table 2.
Let us quickly review these results, particularly under the aspect of
compatibility with our new results.

Probably the first size distribution of soft X-ray fluxes from solar
flares was reported for 177 events observed with OSO-3 by 
Hudson et al.~(1969). The cumulative size distribution shows a powerlaw 
slope of $\beta_P \approx 0.85$ and a gradual turnover at the lower end,
which corresponds to $\alpha_P = \beta_P + 1 \approx 1.85$ and is
roughly consistent with SOC theory ($\alpha_P=2$), given the small-number
statistics.

A larger statistics of 3140 flare events was gathered with the Explorer
spacecraft (Drake et al.~1971), 
finding a powerlaw slope of $\alpha_P=1.84\pm0.02$ if the
fit extended to the third-lowest bin, and a flatter value of 
$\alpha_P=1.66\pm0.02$ if the fit extended to the second-lowest bin.
Clearly, the turnover at low values flattens the powerlaw slope as
expected, a systematic error that is not included in the error bars.
The same study finds also a powerlaw slope of $\alpha_E=1.44\pm0.01$
for the fluence, and thus is largely consistent with SOC theory
($\alpha_P=2.0, \alpha_E=1.5$). 

Shimizu (1995) sampled active region brightenings with Yohkoh, which
we consider as ``miniature flares''. Measuring the peak soft X-ray
intensities averaged over $8\times8$ macropixels, he found a powerlaw
slope of $\alpha_P=1.80$, and $\alpha_P=1.68$ for $16\times16$ macropixels,
while the powerlaw of energies was found to be $\alpha_E \approx 1.5-1.6$,
which is largely consistent with SOC theory ($\alpha_P=2.0, \alpha_E=1.5$). 
Another study on Yohkoh active region brigthenings reported a less accurate
value of $\alpha_P=1.7\pm0.4$ (Shimojo and Shibata 1999), due to the 
relatively small sample of 92 events. 

Lee et al.~(1995) reported powerlaw slopes of $\alpha_P=1.86$ for
soft X-ray peak fluxes observed with BCS/SMM, and a slope of $\alpha_P=1.86$
for a sample of 4356 GOES events. A similar value of $\alpha_P=1.88\pm0.21$
was reported by Feldman et al.~(1997), all within about 5\% of the 
theoretically predicted value. 

A larger dataset of GOES flares during the years of
1986-2000 was analyzed by Veronig et al.~(2002a), yielding a powerlaw
index of $\alpha_P=1.98\pm0.08$ after preflare background subtraction,
which is perfectly consistent with SOC theory and the values obtained
in this study ($\alpha_P=1.98\pm0.11$). In a related study with 49,409
GOES flares, where no preflare background was subtracted, 
a value of $\alpha_P=2.11\pm0.13$
was found (Veronig et al.~2002b). Since the powerlaw range covered
only medium-size to large flares, the background subtraction that is
important for small events did not spoil the powerlaw fit. The powerlaw
fit of soft X-ray flare durations was found to be $\alpha_T=2.93 \pm 0.12$.
This is not inconsistent with our findings, because this powerlaw index
was found to vary in the range of $\alpha_T \approx 2-5$ from the minimum
to the maximum of the solar cycle, which we interpret as flare pile-up bias. 

Very similar values were reported for GOES flares during the period of
1996-2005 (Yashiro et al.~2006), with $\alpha_P=2.16\pm0.03$ for all
flares, and $\alpha_P=1.98\pm0.05$ for flares with CMEs. The complementary
subset of flares without CMEs have a higher value of $\alpha_P=2.52\pm0.03$,
which can be understood as a selection effect that preferentially excludes
the larger flares, and thus is expected to produce a steeper powerlaw slope.
The flare durations showed powerlaw slopes in the range of $\alpha_T
\approx 2.5-3.2$ for the same subsets, which can also be understood in
terms of the flare pile-up bias. 

In summary, all previous measurements are consistent with the theoretically
predicted powerlaw slope of $\alpha_P=2.0$, if we take the uncertainties of
small-number statistics and the turnover at the lower bound of the powerlaw
range into account. The distribution of flare durations was generally
reported to be higher ($\alpha_T \approx 2.5-3.2$) than the theoretically 
predicted value $\alpha_T=2.0$, but can be satisfactorily understood in 
terms of the flare pile-up bias effect.

\subsection{	Coronal Heating by Nanoflares 		}

If a powerlaw distribution of flare energies $E$ is derived, the critical
slope of $\alpha_{E,crit}=2.0$ decides whether the integral of the differential
flare energy distribution diverges at the lower or upper end of the
powerlaw range (Hudson et al.~1991). Our SOC model, which is increasingly
supported by previous observations (Aschwanden 2011b,c) as well as with
the new analysis presented here, predicts a powerlaw slope of
$\alpha_E=1.5$ for total time-integrated energies, which is also consistent
with observed occurrence frequency distributions in hard X-rays
(e.g., $\alpha_E=1.53\pm0.02$; Crosby et al.~1993). Based on this
agreement between the theoretical and observational results there is
neither a theoretical prediction nor observational evidence for a
powerlaw slope $\alpha_E$ steeper than the critical value of 
$\alpha_{E,crit}=2$. In addition, the theoretical SOC model predicts
a powerlaw slope of $\alpha_P=2.0$ for peak energy dissipation rates,
which requires an energy slope of $\alpha_E=1.5$ to be consistent
with SOC theory. This has also been confirmed in this study, which 
further corroborates the theoretical SOC model. Based on this argument 
we conclude that nanoflares have only a negligible contribution to 
coronal heating.

\subsection{	Physical Aspects of the SOC Model 	}

What do we learn about the physics of solar flares by applying a
statistical SOC model? The classical cellular automaton model mimics
the complex spatial and temporal patterns that result in a nonlinear
energy dissipation process, without involving a physical mechanism.
However, it is thought that a solar flare consists of a fragmented
energy release process (Benz 1985), which is manifested in a highly
intermittent time structure of decimetric radio emission or nonthermal
hard X-ray emission (with time structures down to milliseconds).
Numerical MHD simulations can reproduce such a intermittent
time evolution in chain reactions of tearing-mode instabilities and
magnetic island formation (e.g., Sturrock 1966; LeBoef et al. 1982; 
Tajima et al. 1987; Kliem 1990, 1995; Karpen et al.~1995; Drake et 
al. 2006a,b), leading to fractal current sheets (Shibata and Tanuma 2001).
What the SOC model tells us are the scaling laws between the energy
dissipation rate, total dissipated energy, spatial scales, and flare 
durations, since the power indexes of the correlated parameters are
linked to the powerlaw distributions of these parameters. While
SOC theory provides the statistical framework of basic spatio-temporal 
parameters (such as time scales, length scales, and fractal volume scaling),
the physics comes into the problem by connecting the observables
(i.e., flux intensities at different wavelengths) to the geometric 
SOC parameters, which involves physical modeling in terms of electron
densities, electron temperatures, thermal, and nonthermal energies.

In this study we found that the powerlaw slopes are consistent with the
assumption that the emitted soft X-ray flux during a solar
flare is proportional to the fractal flare volume (Eq.~9), 
i.e., $f_{sxr}(t) \propto f(t) \propto V_S(t)$. Given the fact that
soft X-ray emission consists of thermal free-free emission, which in the
optically-thin limit has a proportionality of the emission measure
to the squared density in a volume element cell $dV$,
\begin{equation}
	f_{sxr} \propto EM_{sxr} \propto \int n_e^2 dV \ ,
\end{equation}
the emissivity per volume cell is expected to depend on the electron
density. Physical modeling is required to quantify the soft X-ray emissivity 
and emission measure
per volume element in a given wavelength range as a function of the volumetric 
heating rate, conductive, and radiative loss rate 
(e.g., Metcalf and Fisher 1995; Phillips and Feldman 1995; 
Garcia 1998, 2001; Aschwanden and Alexander 2001; Battaglia et al.~2005;
Sylwester et al.~1995; Veronig and Brown 2004; Veronig et al.~2005;
Yamamoto and Sakurai 2010). 
In summary, SOC theory represents an important statistical tool 
to link physical modeling to observed correlations (scaling laws)
and size distributions. 

\subsection{	Soft X-ray versus Hard X-ray Flare Statistics		}

A compilation of hard X-ray flare statistics over the last three solar cycles
has been conducted in a recent study (Aschwanden 2011b and references therein).
The following powerlaw slopes have been found for the hard X-ray parameters:
$\alpha_P=1.73\pm0.07$ for the hard X-ray peak flux $P$,
$\alpha_E=1.62\pm0.12$ for the total time-integrated hard X-ray flux (or fluence) $E$,
$\alpha_T=1.99\pm0.35$ for the hard X-ray flare durations $T$. 
These results agree remarkably well with ouf FD-SOC model, which predicts
$\alpha_P=5/3 \approx 1.67$, $\alpha_E=3/2=1.50$, and $\alpha_T=2.00$
for the 3D case (Aschwanden 2012; Table 1 therein). Comparing hard X-ray
with soft X-ray flare statistics, we have to be aware that the peak fluxes
in these two wavelengths are not equivalent. The peak of the hard X-ray flux
light curve $f_{hxr}(t)$ represents a maximum fluctuation of the instantaneous 
energy dissipation rate (if the temporal fluctuations are fully resolved in time), 
while the peak flux in soft X-rays represents the maximum value of the smoothed 
energy dissipation rate, because the intermittent and spiky instantaneous energy 
dissipation rate is convolved with a cooling time, which has a smoothing effect 
on the light curve, as demonstrated in Fig.~7. Therefore, our FD-SOC model
predicts a steeper slope $\alpha_F=2.0$ for the soft X-ray peak flux distribution
than the slope $\alpha_P=1.67$ for the hard X-ray peak flux distribution, which
is indeed confirmed with the observed data.

In an earlier study (Lee et al. 1995) it was suspected that the size distribution
of the soft peak flux (with powerlaw slope $\alpha_F$) should correspond to the 
size distribution of the (time-integrated) hard X-ray fluence $\alpha_E$, according
to the Neupert effect, which predicts
\begin{equation}
	f_{sxr}(t) = \int_0^t f_{hxr}(t') dt' \ ,
\end{equation}   
but the data were found to contract this expectation (Lee et al.~1995). 
This failure corroborates that the integral formulation (Eq.~21) of the
Neupert effect is an oversimplified approximation and does not
hold on a statistical basis. A more accurate approximation is the 
formulation of the Neupert effect in terms of a convolution of the 
hard X-ray flux with a finite
e-folding cooling time, as expressed with Eq.~(11). Such a relationship is
consistent with the predictions of our FD-SOC model, which predicts 
a powerlaw slope of $\alpha_F=2.0$ for the soft X-ray peak flux and
a powerlaw slope of $\alpha_E=1.5$ for the hard X-ray fluence (while the
Neupert effect would predict both to have an identical value).
The relationship between soft X-ray and hard X-ray fluxes has also been 
quantified in some statistical studies (e.g., Battaglia and Benz 2006; 
McTiernan 2009; Falewicz et al.~2009).

The powerlaw slope $\alpha_P$ of the hard X-ray peak flux was found to vary
sligthly with the solar cycle, between a minimum value of $\alpha_P=1.62\pm0.02$
in the decay phase of a solar cycle to $\alpha_P=1.79\pm0.02$ during the 
minimum or rise of the solar cycle (Aschwanden 2012a). However, this variation
is about of the same order as the differences between the three different 
instruments (SMM, CGRO, RHESSI) and three different flare detection methods
used for the statistics. Since we find that some soft X-ray parameters are
invariant during the solar cycle, while other parameters vary due to the
flare pile-up bias, it is not clear at this point whether the variation of
hard X-ray powerlaws is also caused by the flare pile-up bias, or if it
is related to intrinsic variations of the self-organized criticality 
conditions (Aschwanden 2011c). 

\subsection{	Solar Flare Predictability 		}

The occurrence frequency distributions we obtained from the GOES
soft X-ray fluxes are also called probability distribution functions (PDF)
and can be used for statistical flare prediction. The fact that we
established here the invariance of the powerlaw slope, makes the
statistical prediction very easy, because the number of flares observed
or to be predicted depends only on one single time-dependent parameter,
the scaling factor $N_1$,
\begin{equation}
	N(P) dP =  N_1(t) \left( {P \over P_1} \right)^{-\alpha_P} dP = 
		   N_1(t) \left( {P \over P_1} \right)^{-2} dP \ ,
\end{equation}
where $P_1$ is an absolute constant, say $P_1=10^{-6}$ $W m^{-2}$ for the
C-class level, and $N_1(t)$ is the number of flares observed at this level
in the interval $dP$.
Forecasting the number of flares that are larger than a given threshold
level, are given by the cumulative PDF,
\begin{equation}
	N^{cum}(>P) =  \int_P^{\infty} N_1(t) 
	\left( {P \over P_1} \right)^{-2} dP = 
	N_1(t) \left( {P \over P_1} \right)^{-1} \ .
\end{equation}
The time-varying function $N_1(t)$ can be measured at any level $P_1$
within the powerlaw range. This time-varying function $N_1(t)$ can be
broken down for different active regions and be turned on and off 
individually for each active region in a prediction model, 
depending on their appearance on the
east-side of the Sun or disappearance on the west-side. Thus, the
accuracy of forecasting depends only on the temporal extrapolation of the
flaring rate $N_1(t)$ per active region, which can also be trended
from statistical variations.

Predicting the largest event in history or future is a delicate issue.
In principle we can take the mean of the largest events during each 
solar cycle as a reasonable guess for the next solar cycle, which yields
something like an $X20$ GOES class. The solar
dynamo resets the magnetic field after every 11-year cycle, so that no
nonpotential magnetic field can be stored on longer terms, potentially
leading to larger giant flares. What the SOC theory tells us in addition
is the scaling of the flare energy with the length scale, which simply
scales with the fractal volume, $E \propto V_S \propto L^{D_S}$.
Once the size scale $L$ is measured for the largest flares, we can scale
the maximum possible flare energy by the maximum scale size $L_{max}$, 
which probably corresponds to the maximum active region size $L_{AR}$.
Active regions grow and wane in size, $L_{AR}(t)$, which can be used
to estimate the maximum flare rate $N_1(t)$ per active region, and this
way be used in the prediction of the cumulative PDF of flares above 
some size (Eq.~23).

\section{		CONCLUSIONS			}

In order to obtain a deeper physical understanding of nonlinear dissipative
systems governed by self-organized criticality (SOC) we need large statistics 
of SOC parameters to accurately quantify their statistical probability
distribution functions (PSF) and the scaling laws between SOC parameters.
For this purpose we analyzed the largest available, uniformly sampled dataset
of solar flare light curves from the GOES spacecraft series over the last 37
years (1974-2012). Analyzing this dataset and applying the fractal-diffusive
self-organized criticality model (FD-SOC) we arrived at the following 
major conclusions:

\begin{enumerate}
\item{With an automated flare detection algorithm applied to the GOES
	1-8 \ang\ light curves we detect a total of 338,661 flare events
	during the epoch of 1975-2011 that includes all full years with GOES
	records. Our algorithm detects about 5 times more solar flares (with 
	a noise level at the GOES A2-class level) than the official GOES flare 
	catalog issued by NOAA, which covers the epoch of 1991-2012.
	The duty cycle of GOES is found to be 94\%$\pm$4\% during the years
	1978-2011.} 
\item{The occurrence frequency distributions of GOES flare parameters can be
	characterized by a powerlaw function with a gradual turnover at the
	lower end, which can be characterized and fitted with the function 
	$N(x) = N_1 (1 + x/x_1)^{-\alpha}$,
	in order to obtain an accurate powerlaw slope corrected for the turnover.
	We find the following mean powerlaw slopes: $\alpha_F=1.98\pm0.11$ for
	the soft X-ray peak flux $F$, $\alpha_T=2.97 \pm 0.71$ for the soft X-ray
	rise time $T$, and $\alpha_{df/dt}=2.01\pm0.12$ for the steepest
	time derivative during the soft X-ray rise time.}
\item{The powerlaw slopes of the (background-subtracted) peak flux $F$ and 
	of the time derivative $(df/dt)$ are found to be invariant during the
	last three solar cycles with a variation of less than 5\%.
	The powerlaw slope of the soft X-ray rise time $T$ and the turnover values
	($F_1, df/dt_1, T_1$) are all found to vary systematically with 
	the solar cycle and can be modeled with the flare pile-up bias effect,
	which causes a loss of small flares and an underestimate of the flare
	duration (defined by the rise time here) during times of high flare rates.
	The flare rate of years during the solar minimum, however, are not
	affected by the flare pile-up bias, during which we measure a powerlaw
	slope of $\alpha_T=2.02\pm0.04$ for the soft X-ray rise time $T$.} 
\item{The fractal-diffusive self-organized criticality (FD-SOC) model
	(Aschwanden 2011a) predicts a fractal dimension of $D_3=2.0$ for
	flare volumes and powerlaw distributions with slopes of $\alpha_F=2.0$
	for the (smoothed) peak flux $F$ and $\alpha_T=2.0$ for the flare
	duration, which are both consistent with the observations (taking
	the flare pile-up effect into account). Most previous studies 
	are also consistent with our results and the FD-SOC model (if we
	take flare pile-up effects into account), except those with small-number 
	statistics or no background subtraction.}
\item{The consistency of the observations with the theoretical (FD-SOC) model
	implies also a powerlaw distribution for the time-integrated dissipated
	flare energy with a powerlaw slope of $\alpha_E=1.5$, which rules out
	that nanoflares contribute significantly to coronal heating (which would
	require a powerlaw slope of $\alpha_E > \alpha_{crit}=2.0$.)}
\item{The theoretical FD-SOC model predicts the powerlaw distributions of
	the instantaneous energy dissipation rate $F$, the peak dissipation rate
	$P$, the time-integrated energy release volume $E$, the time duration $T$,
	the fractal dimension $D_S$ of the avalanches volume $V_S$, and scaling 
	laws between these spatial and temporal parameters of SOC avalanches.
	The prediction of the size distributions of observables, such as the
	soft X-ray flux $F_{sxr}$ or hard X-ray flux $F_{hxr}$ requires a
	physical model of the emissivity per volume element, which entails
	thermal and non-thermal emission. For soft X-ray data we find that the
	soft X-ray intensity $F_{SXR}$ is proportional to the instantaneous
	energy dissipation rate $F$, which implies a balance between the
	plasma heating rate and the radiative loss rate.}
\item{The hard X-ray peak fluxes have a powerlaw slope $\alpha_P \approx 1.7$ 
	that is different from the powerlaw slope $\alpha_F=2.0$ of the soft 
	X-ray peak flux, but both values are consistent with the FD-SOC theory.
	The difference can be understood by the fact that the hard X-ray peak flux
	represents the most extreme fluctuation of the energy dissipation rate
	during a flare, while the soft X-ray peak flux represents a convolution
	function with an e-folding cooling time, which smoothes out the
	fluctuations of the intermittent energy dissipation rate. The integral
	formulation of the Neupert effect appears to be an oversimplification
	that is not consistent with the data, but a formulation in terms of
	a convolution of the hard X-ray flux with an e-folding cooling time
	seems to be more realistic and is consistent with the FD-SOC model.} 
\item{The invariance of the soft X-ray peak flux powerlaw slope ($\alpha_F=2$)
	during the solar cycles simplifies the statistical prediction of 
	solar flares, since the statistical expectation value is then 
	proportional for every flux level. The cumulative occurence frequency
	distribution has a slope of $\beta_F=\alpha_F-1=1$, which states that
	the probability of a flare greater than a flux level $F$ drops
	reciprocally with the flux level $F$. An X-class flare is 100 times
	less likely than a C-class flare. The largest flares in history or
	future are unlikely to surpass the largest flares observed hitherto, 
	based on the fact that the FD-SOC model predicts a dependence of the released
	energy on the flare size, which is limited by the largest system size,
	i.e., the largest active region in the case of solar flares.}
\end{enumerate}

If the reader goes back to the eight questions at the beginning of the
Introduction, he/she will find that we arrived at a quantitative answer
to almost all questions. Nevertheless, there are more open questions:
How can we understand the solar cycle variation of the hard X-ray powerlaw slope?
What is the physical scaling between the (fractal) avalanche volume and the
emissivity in different wavelengths (soft X-rays, hard X-rays, gamma-rays, EUV)?
How do the scaling laws between different forms of flare energies relate to
each other (thermal, non-thermal, kinetic, magnetic energy)?
Does a more extended flare energy distribution spanning over 8 orders of
magnitude from the largest flare to the smallest detectable nanoflares
exhibit a single powerlaw slope? Can we statistically verify the fractal 
spatial geometry and scaling laws with imaging observations of solar flares?
Future studies with multi-wavelength data from AIA/SDO are expected to shed
more light on these problems.

\acknowledgements
The first author acknowledges helpful discussions with Karel Schrijver
and Greg Slater. We thank the data services at NOAA, NASA/GSFC Solar 
Data Analysis Center (SDAC), SSW support, and the individuals who
contributed to the GOES/SSW database, in particular Joe Gurman, Kim Tolbert,
Dominic Zarro, Richard Schwartz, and Mons Morrison.


\clearpage


\begin{deluxetable}{rrcrrrr}
 \tabletypesize{\normalsize}
\tablecaption{Powerlaw slopes of occurrence frequency distributions 
detected with GOES per year.}
\tablewidth{0pt}
\tablehead{
\colhead{Year}&
\colhead{Number}&
\colhead{Number of}&
\colhead{Duty}&
\colhead{Powerlaw slope}&
\colhead{Powerlaw slope}&
\colhead{Powerlaw slope}\\
\colhead{}&
\colhead{of events}&
\colhead{NOAA events}&
\colhead{cycle}&
\colhead{peak flux $\alpha_F$}&
\colhead{rise time $\alpha_T$}&
\colhead{derivative $\alpha_{df/dt}$}}
\startdata
Mean &   9153 & 1804 &  0.922 & 1.98$\pm$0.11 & 2.97$\pm$0.71 & 2.01$\pm$0.12 \\
     &        &      &        &               &               &               \\
1975 &   2865 & .... &  0.794 & 2.04$\pm$0.10 & 2.24$\pm$0.04 & 1.98$\pm$0.08 \\ 
1976 &   1827 & .... &  0.670 & 1.82$\pm$0.07 & 2.13$\pm$0.02 & 1.89$\pm$0.06 \\ 
1977 &   5543 & .... &  0.818 & 1.91$\pm$0.04 & 2.43$\pm$0.06 & 1.96$\pm$0.05 \\ 
1978 &  16897 & .... &  0.923 & 1.92$\pm$0.01 & 2.76$\pm$0.07 & 1.95$\pm$0.04 \\ 
1979 &  18797 & .... &  0.910 & 2.00$\pm$0.03 & 3.18$\pm$0.10 & 2.04$\pm$0.04 \\ 
1980 &  15340 & .... &  0.843 & 2.10$\pm$0.04 & 2.72$\pm$0.05 & 2.13$\pm$0.03 \\ 
1981 &  15527 & .... &  0.871 & 2.07$\pm$0.02 & 3.22$\pm$0.09 & 2.05$\pm$0.01 \\ 
1982 &  15963 & .... &  0.914 & 1.96$\pm$0.04 & 2.89$\pm$0.03 & 1.96$\pm$0.02 \\ 
1983 &   9566 & .... &  0.897 & 2.04$\pm$0.03 & 2.72$\pm$0.03 & 2.01$\pm$0.03 \\ 
1984 &   6894 & .... &  0.873 & 1.95$\pm$0.01 & 2.77$\pm$0.03 & 1.99$\pm$0.05 \\ 
1985 &   2776 & .... &  0.904 & 1.98$\pm$0.07 & 2.37$\pm$0.05 & 1.87$\pm$0.06 \\ 
1986 &   2337 & .... &  0.900 & 1.87$\pm$0.04 & 2.27$\pm$0.01 & 2.03$\pm$0.09 \\ 
1987 &   4863 & .... &  0.865 & 2.11$\pm$0.07 & 2.67$\pm$0.05 & 1.98$\pm$0.06 \\ 
1988 &  13030 & .... &  0.947 & 2.07$\pm$0.03 & 3.24$\pm$0.11 & 2.10$\pm$0.03 \\ 
1989 &  17535 & .... &  0.938 & 2.00$\pm$0.04 & 3.81$\pm$0.10 & 2.03$\pm$0.04 \\ 
1990 &  15280 & .... &  0.944 & 2.13$\pm$0.02 & 3.44$\pm$0.55 & 2.11$\pm$0.03 \\ 
1991 &  16860 & 2124 &  0.943 & 2.02$\pm$0.03 & 3.29$\pm$0.08 & 2.02$\pm$0.01 \\ 
1992 &  12488 & 4434 &  0.933 & 1.96$\pm$0.04 & 3.17$\pm$0.09 & 2.01$\pm$0.02 \\ 
1993 &   9386 & 3181 &  0.902 & 2.02$\pm$0.02 & 3.41$\pm$0.21 & 2.11$\pm$0.05 \\ 
1994 &   4844 & 1694 &  0.958 & 2.03$\pm$0.03 & 2.64$\pm$0.03 & 2.10$\pm$0.02 \\ 
1995 &   2650 & 1067 &  0.847 & 1.99$\pm$0.02 & 2.58$\pm$0.05 & 1.96$\pm$0.05 \\ 
1996 &   1288 &  551 &  0.937 & 1.93$\pm$0.06 & 2.20$\pm$0.03 & 2.00$\pm$0.05 \\ 
1997 &   3776 & 1148 &  0.973 & 1.99$\pm$0.04 & 2.51$\pm$0.05 & 1.98$\pm$0.06 \\ 
1998 &  11172 & 2256 &  0.978 & 1.96$\pm$0.04 & 3.26$\pm$0.12 & 2.04$\pm$0.02 \\ 
1999 &  14685 & 2377 &  0.976 & 2.12$\pm$0.04 & 3.02$\pm$0.13 & 2.17$\pm$0.05 \\ 
2000 &  16156 & 2661 &  0.972 & 2.04$\pm$0.03 & 4.61$\pm$0.15 & 2.10$\pm$0.03 \\ 
2001 &  16022 & 2710 &  0.979 & 1.97$\pm$0.02 & 3.51$\pm$0.62 & 2.03$\pm$0.06 \\ 
2002 &  16167 & 2670 &  0.965 & 2.07$\pm$0.05 & 5.18$\pm$0.55 & 2.15$\pm$0.04 \\ 
2003 &  12283 & 2395 &  0.978 & 1.85$\pm$0.02 & 3.82$\pm$0.19 & 1.85$\pm$0.03 \\ 
2004 &   8956 & 2368 &  0.962 & 1.88$\pm$0.04 & 3.12$\pm$0.11 & 1.95$\pm$0.03 \\ 
2005 &   6706 & 2029 &  0.970 & 1.77$\pm$0.01 & 3.19$\pm$0.16 & 1.83$\pm$0.02 \\ 
2006 &   3100 & 1200 &  0.983 & 1.89$\pm$0.06 & 2.43$\pm$0.03 & 1.92$\pm$0.04 \\ 
2007 &   1434 &  651 &  0.947 & 1.88$\pm$0.05 & 2.17$\pm$0.02 & 2.00$\pm$0.05 \\ 
2008 &    186 &   88 &  0.981 & 1.69$\pm$0.07 & 1.75$\pm$0.05 & 1.72$\pm$0.04 \\ 
2009 &    528 &  245 &  0.979 & 2.31$\pm$0.14 & 1.99$\pm$0.05 & 2.48$\pm$0.06 \\ 
2010 &   3662 & 1261 &  0.965 & 1.92$\pm$0.06 & 3.76$\pm$0.69 & 1.91$\pm$0.06 \\ 
2011 &  11272 & 2171 &  0.973 & 2.00$\pm$0.03 & 3.29$\pm$0.41 & 2.02$\pm$0.03 \\ 
\enddata
\end{deluxetable}


\begin{deluxetable}{llllrll}
 \tabletypesize{\normalsize}
\tablecaption{Frequency distributions measured from solar flares in 
soft X-rays. References: 
$^1)$ Hudson et al. (1969);  
$^2)$ Drake et al. (1971);   
$^3)$ Shimizu (1995);    
$^4)$ Lee et al. (1995);  
$^5)$ Feldman et al. (1997); 
$^6)$ Shimojo and Shibata (1999); 
$^7)$ Veronig et al. (2002b); 
$^8)$ Veronig et al.~(2002a);
$^9)$ Yashiro et al.~(2006);
$^{10})$ This work;  
$^{11})$ Aschwanden (2012);   
Data analysis with no background subtraction are marked with 
the symbol $^*$.}
\tablewidth{0pt}
\tablehead{
\colhead{Powerlaw}&
\colhead{Powerlaw}&
\colhead{Powerlaw}&
\colhead{log size}&
\colhead{Number}&
\colhead{Instrument}&
\colhead{Reference}\\
\colhead{slope of}&
\colhead{slope of}&
\colhead{slope of}&
\colhead{range}&
\colhead{of events}&
\colhead{}&
\colhead{}\\
\colhead{peak flux}&
\colhead{total fluence}&
\colhead{durations}&
\colhead{}&
\colhead{}&
\colhead{}&
\colhead{}\\
\colhead{$\alpha_P$}&
\colhead{$\alpha_E$}&
\colhead{$\alpha_T$}&
\colhead{}&
\colhead{}&
\colhead{}&
\colhead{}}
\startdata
$\approx 1.85$  &               &           &1       & $\approx 177$ &OSO-3&1)\\
1.84 (1.66)     &1.44           &           &2          &   3,140 &Explorer&2)\\
1.80 (1.68)     &1.5-1.6        &           &2          &     291 &Yohkoh &3)\\
1.79            &               &           &2          &       ? &SMM/BCS&4)\\
1.86            &               &           &2          &   4,356 &GOES &4)\\
1.88$\pm$0.21   &               &           &3          &   1,054 &GOES &5)\\
1.7$\pm$0.4     &               &           &2          &      92 &Yohkoh &6)\\
$1.98\pm0.08$   &$1.89\pm0.10$  &           &3          &       ? &GOES &7,8)\\
$2.11\pm0.13^*$ &$2.03\pm0.09^*$ &$2.93\pm0.12^*$ &3    &  49,409 &GOES &8)\\
$2.16\pm0.03^*$ &$2.01\pm0.03^*$ &$2.87\pm0.09^*$ &3    &   5,890 &GOES &9)\\
$1.98\pm0.11$   &               &2.01$\pm$0.12&4        & 327,389 &GOES &10)\\
$2.0$		&$1.5$		&$2.0$       &          & -       &Theory&11)\\
\enddata
\end{deluxetable}

\clearpage

\begin{figure}
\plotone{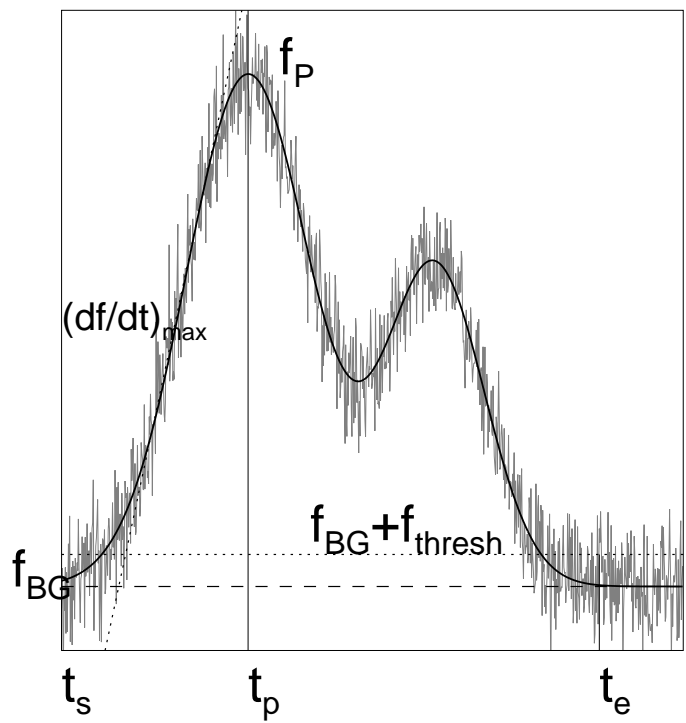}
\caption{Schematic of flare parameters: flare start time $t_s$,
flare peak time $t_p$, flare end time $t_e$, preflare background $f_{BG}$,
flux threshold $f_{thresh}$, flare peak flux $f_P$, and
maximum flux-time derivative $(df/dt)_{max}$. The curve with thick linestyle
represents the rebinned and smoothed light curve which defines the local
flux maxima and minima.}
\end{figure}

\begin{figure}
\plotone{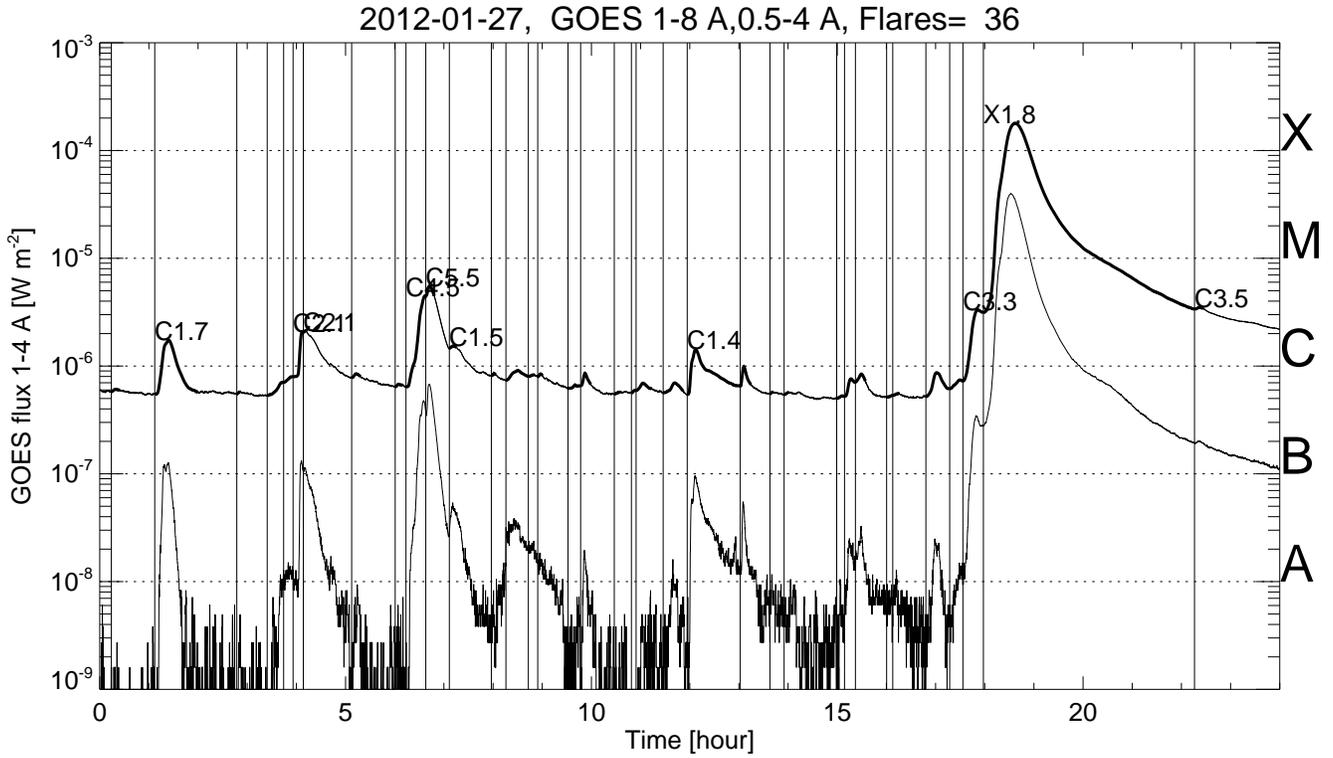}
\caption{Example of automated flare detection of GOES observations during 
the day of 2012-Jan-27, which featured an X1.8 flare. A number of 36 flares
are detected down to the GOES B5-class level (with the starting times $t_s$
marked with vertical bars). Only flares with a GOES class of $>C1.0$ are 
labeled.
The upper light curve is the 1-4 \ang\ wavelength range, which is used for
flare detection, while the lower curve shows the 0.5-8 \ang\ data.}
\end{figure}

\begin{figure}
\plotone{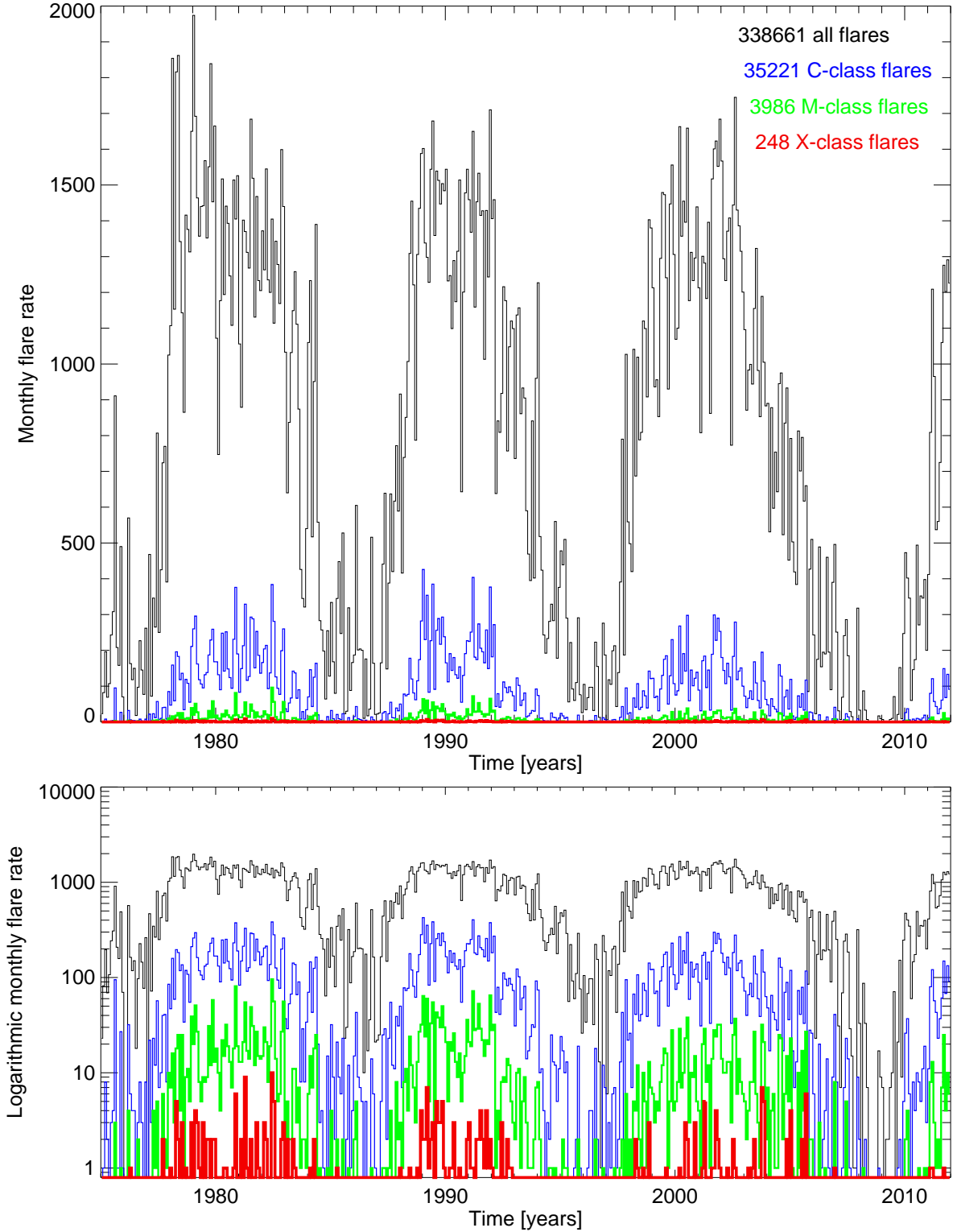}
\caption{Monthly flare rate during 1975-2011 on a linear scale (top panel)
and on a logarithmic scale (bottom panel). The rate of C (blue), M (green), 
and X-class flares (red) are shown in colors. Note the proportionality of
detected flares in different magnitudes.} 
\end{figure}

\begin{figure}
\plotone{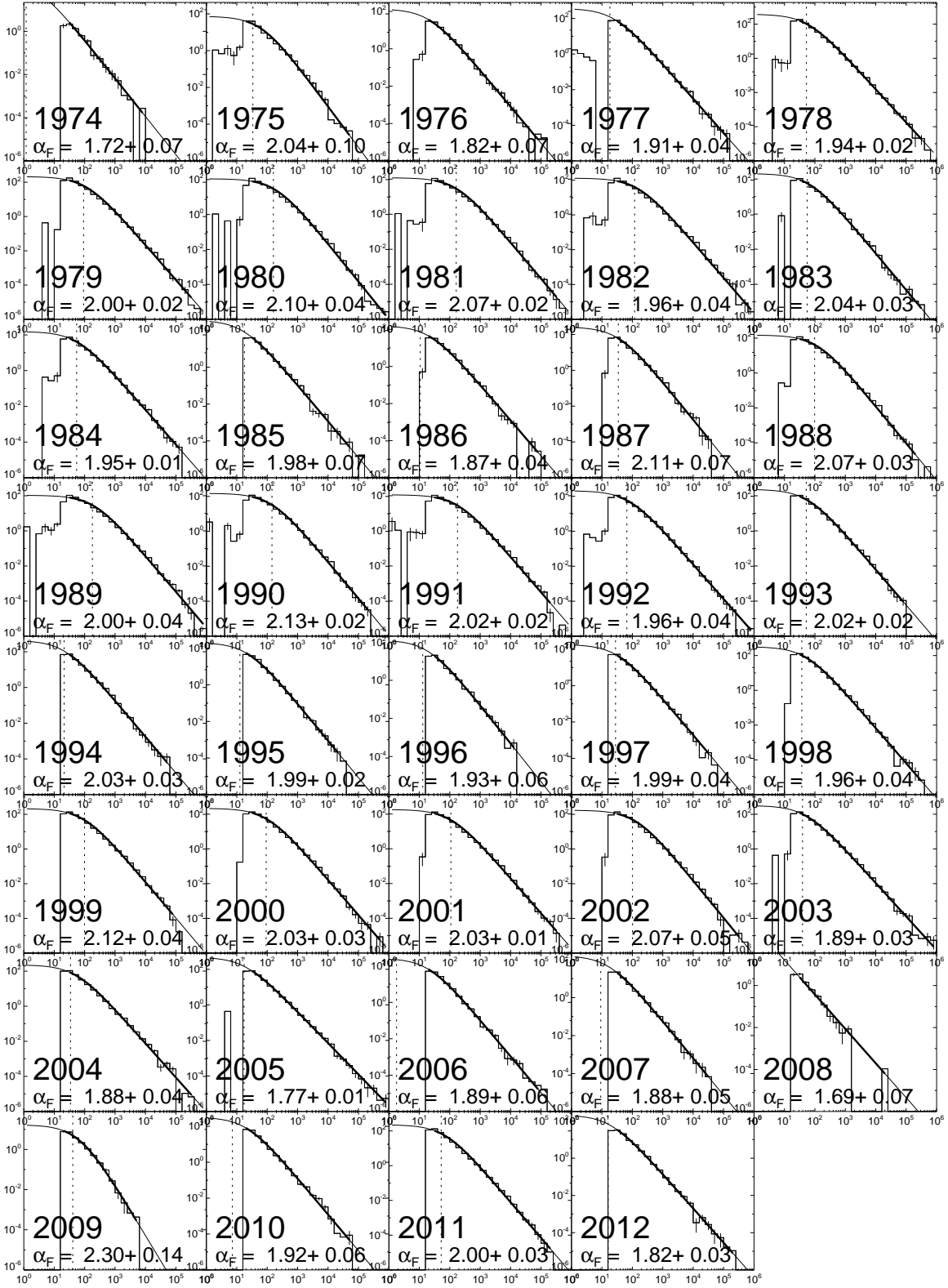}
\caption{Occurrence frequency distributions of GOES 1-4 \ang\ peak fluxes
of solar flares detected by year. The powerlaw slope $\alpha_F$ is indicated
for each year. Incomplete years are the first (1974 Nov-Dec) and the last
one (2012 Jan-Mar).}
\end{figure}

\begin{figure}
\plotone{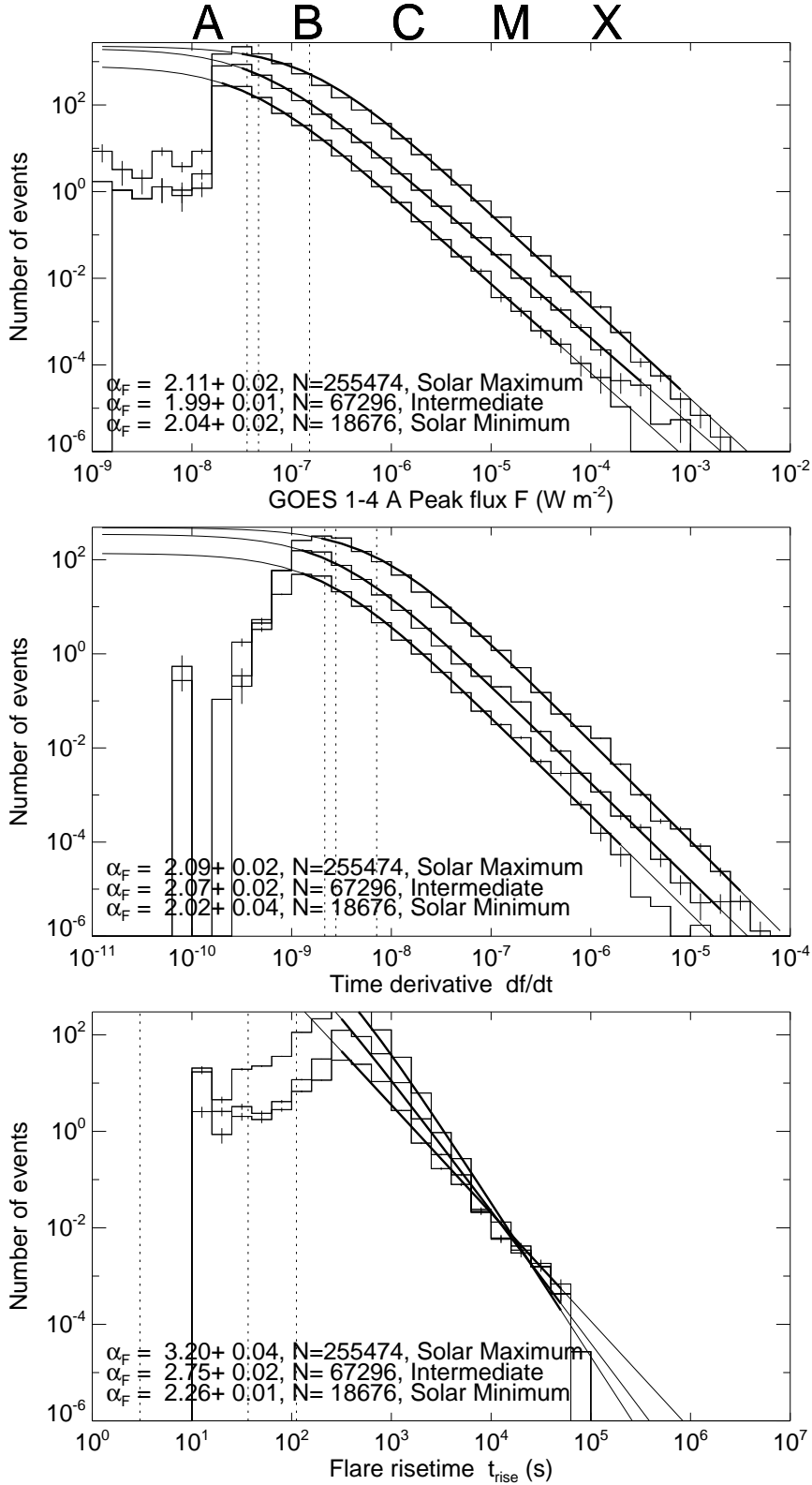}
\caption{Occurrence frequency distribution of the GOES 1-4 \ang\ peak fluxes
(top panel), the time derivative $(df/dt)_{max}$ (middle panel), and
flare rise times $t_{rise}$  
during the solar cycle maximum (years with $N_{ev} \ge 10000$ events),
the solar cycle minimum (years with $N_{ev} < 3000$ events), and for
intermediate years. Note the invariance of the powerlaw slope for the
peak fluxes (top panel) and time derivatives (middle panel).}
\end{figure}

\begin{figure}
\plotone{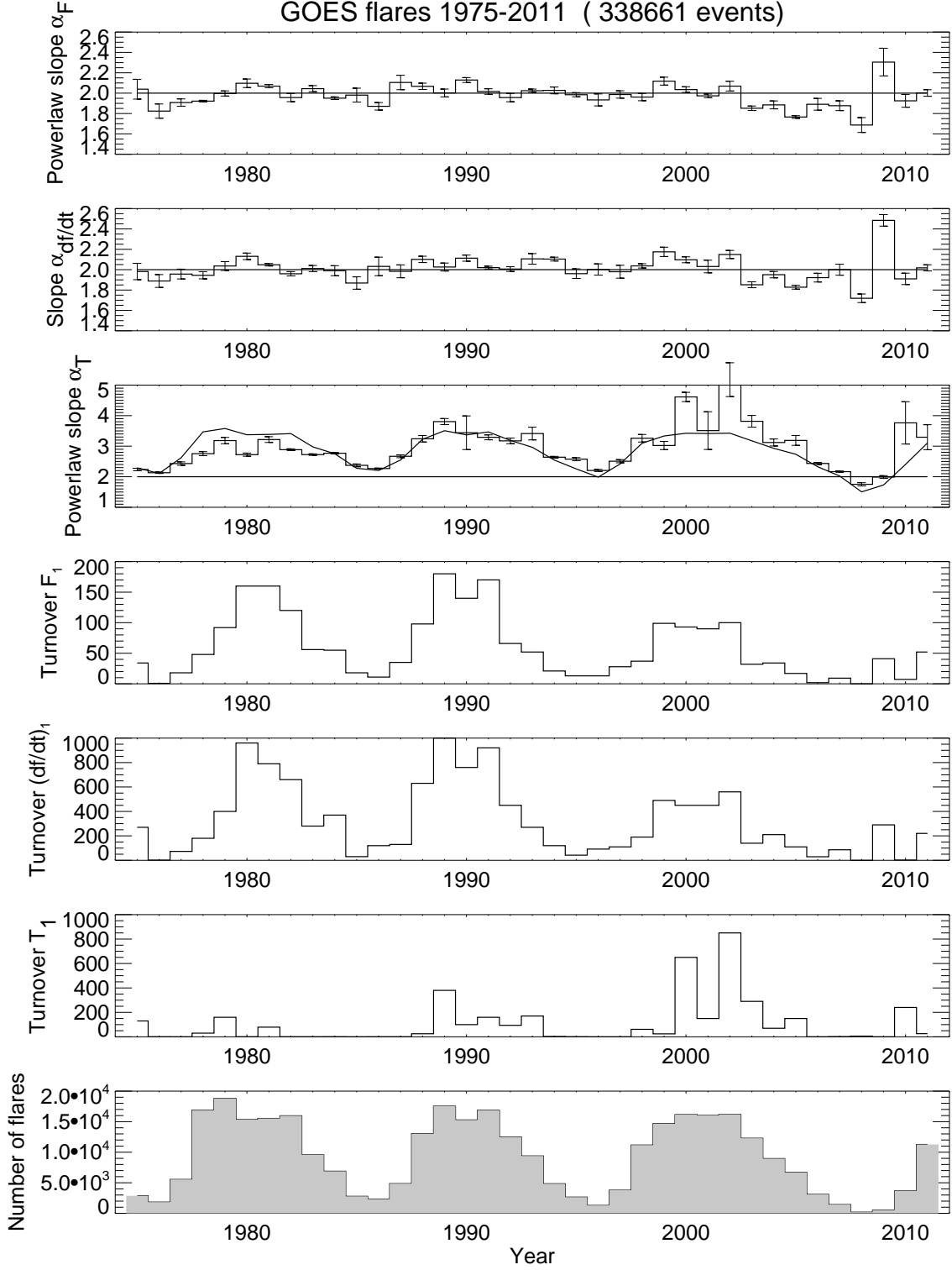}
\caption{Variation of powerlaw slopes $\alpha_F(t)$, flux-time derivative
$\alpha_{df/dt}(t)$, rise time $\alpha_T(t)$, turnover flux value $F_1$
(in units of $10^{-9}$ W m$^{-2}$),
turnover value of time derivative $(df/dt)_1$, 
(in units of $3 \cdot 10^{-11}$ W m$^{-2}$ s$^{-1}$), 
turnover value of rise time $T_1(t)$ (in units of $s$), 
and flare rate $N_{flare}(t)$ during the three last solar cycles.
The theoretically predicted values are marked with thick solid lines).}
\end{figure}

\clearpage

\begin{figure}
\plotone{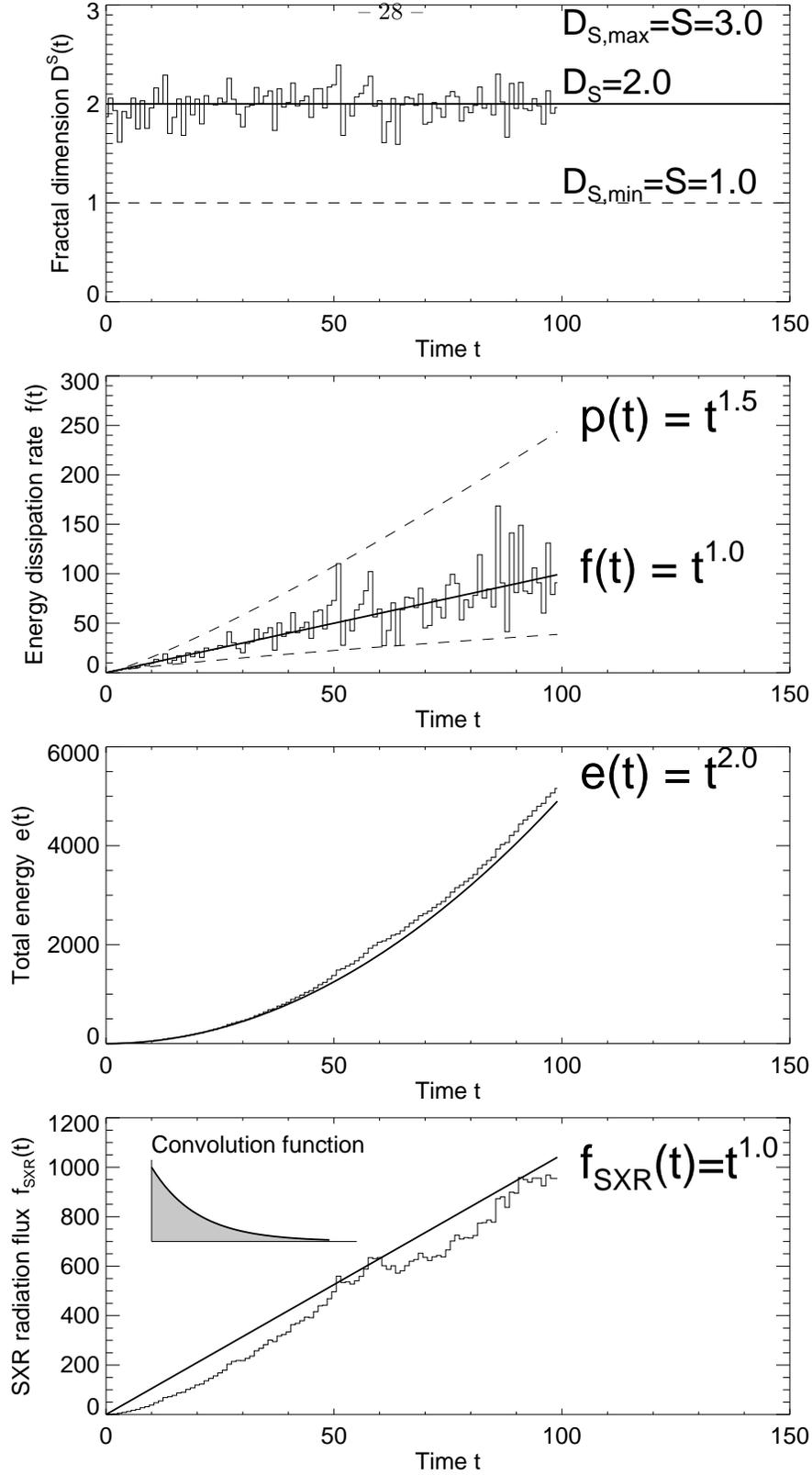}
\caption{Simulation of the fractal-diffusive SOC model for an
Euclidean dimension $S=3$, showing the time evolution of the fractal
dimension $D_{S}(t)$ (top panel), the instantaneous energy dissipation
rate $f(t)$ and peak energy dissipation rate $p(t)$ (second panel),
the total time-integrated dissipated energy $e(t)$ (third panel), 
and the soft X-ray
time profile $f_{sxr}(t)$ (bottom panel), which results from the
convolution of the instantaneous energy dissipation rate $f(t)$ (second
panel) with an exponential decay function with an e-folding time of
$\tau_{decay}=20$ (shown in insert of bottom panel).}
\end{figure}

\end{document}